\documentclass[12pt]{iopart}

\usepackage{cite}

\usepackage{graphicx}
\usepackage[german, english]{babel}
\usepackage{color}

\begin{document}

\title{Resonant five-body recombination in an ultracold gas of bosonic atoms}


\author{Alessandro Zenesini$^{1}$\footnote{New address: Institut f\"ur Quantenoptik, Leibniz Universit\"at Hannover, 30167 Hannover, Germany.},
Bo Huang$^{1}$,
Martin Berninger$^{1}$,
Stefan Besler$^{1}$,
Hanns-Christoph N\"{a}gerl$^{1}$,
Francesca Ferlaino$^{1}$,
Rudolf Grimm$^{1, 2}$,
Chris H Greene$^{3}$\footnote{New address: Department of Physics, Purdue University, West Lafayette, IN 47907, USA},
Javier von Stecher$^{3, 4}$}

\address{$^1$ Institut f\"ur Experimentalphysik und Zentrum f\"ur Quantenphysik, Universit\"at Innsbruck, 6020 Innsbruck, Austria}
\address{$^2$ Institut f\"ur Quantenoptik und Quanteninformation, \"Osterreichische Akademie der Wissenschaften, 6020 Innsbruck, Austria}
\address{$^3$ Department of Physics and JILA, University of Colorado, Boulder, CO 80309, USA}
\address{$^4$ Tech-X Corporation, Boulder, CO 80303, USA}

\ead{alessandro.zenesini@ultracold.at}

\begin{abstract}
We combine theory and experiment to investigate five-body recombination in an ultracold gas of atomic cesium at negative scattering length. A refined theoretical model, in combination with extensive laboratory tunability of the interatomic interactions, enables the five-body resonant recombination rate to be calculated and measured. The position of the new observed recombination feature agrees with a recent theoretical prediction and supports the prediction of a family of universal cluster states at negative $a$ that are tied to an Efimov trimer.
\end{abstract}


\maketitle

\section{Introduction}

Few-body physics with ultracold atoms has emerged as a new research field combining concepts from atomic, nuclear and condensed-matter physics. A growing number of experimental and theoretical studies have been focused on both the fundamentals of few-body phenomena \cite{Braaten2006uif} and the connections with many-body systems \cite{Ferlaino2010fyo, greene2010uif}. The cornerstone of recent experimental advances is the control of interactions in an ultracold atomic gas, offered by magnetically tuned Feshbach resonances \cite{Chin2010fri}. In particular, the tunable $s$-wave scattering length $a$ allows to access the regime of resonant two-body interactions. Here, the system is governed by universal behavior, independent of the short-range details of the interaction potential. The paradigm of universality is Efimov's solution to the problem of three resonantly interacting particles \cite{Efimov1970ela}. Once the intimate connection between Efimov states and three-body recombination has been established \cite{Nielsen1999ler, Esry1999rot, Braaten2006uif}, resonant loss features became the fingerprint of Efimov physics in experimental studies with ultracold atoms \cite{Kraemer2006efe, Ottenstein2008cso, Huckans2009tbr, Knoop2009ooa, Zaccanti2009ooa, Barontini2009ooh, Gross2009oou, Nakajima2010nea, Gross2010nsi, Lompe2010ads, Nakajima2011moa, Pollack2009uit, Ferlaino2011eri}.

\begin{figure}
\begin{center}
\includegraphics[width=.65\textwidth] {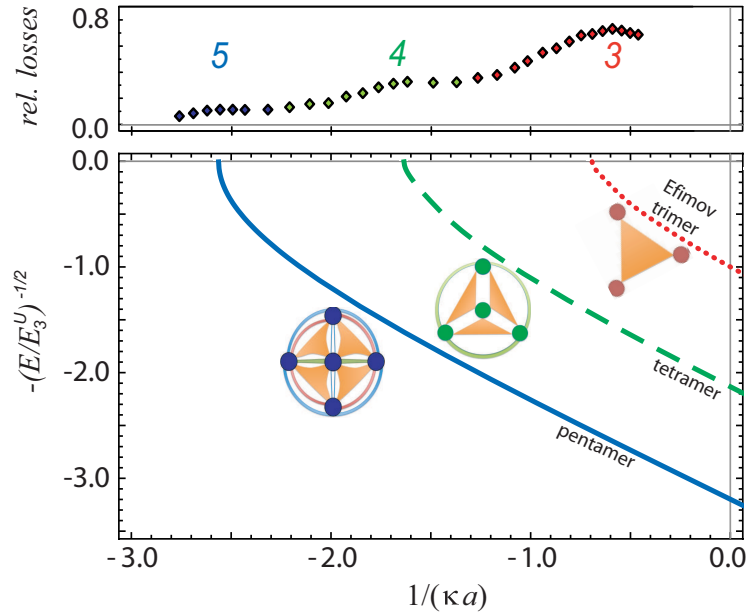}
\caption{$N$-body scenario in the region of negative two-body scattering length $a$. The lower panel shows the $N$-body binding energies as functions of the inverse scattering length. $E^U_3=(\hbar \kappa)^2/m$ is the trimer binding energy for resonant interaction. The dotted, dashed and solid lines refer to the ground states of three-, four- and five-body cluster, respectively. Excited states are not drawn here. The pictorial representations of the three-, four-, and five-body bound state illustrate one of the possible spatial arrangements of atoms, given that an $N$-body cluster state has $N(N-1)(N-2)/6$ possible combinations of the three-body sub-clusters. The upper panel shows the relative losses from the atomic sample for a typical set of measurements, performed with a hold time of 100\,ms for a sample of $10^5$ atoms at 100\,nK in a trap with mean trap frequency of about 30\,Hz.}
\label{Fig.1}
\end{center}
\end{figure}

Advances in three-body physics led to intriguing questions on the generalization of Efimov's scenario to more particles and on the existence and observability of universal few-body cluster states \cite{Blume2000mch,Blume2002foa, Adhikari1981fbe, Naus1987tee, Platter2004fbs, Hanna2006eas}.
$N$-body cluster states known as \textit{Brunnian} states exist in a range of interaction where no ($N-1$)-body weakly bound subsystems are present \cite{Richard1994lot,Yamashita2011bae}. The general connection, however, between cluster states and the three-body Efimov states has remained an open issue. It was soon realized that no ``true'' Efimov states\footnote{The ``true'' Efimov effect refers to the appearance of an infinite number of $N$-body bound states, which have a discrete scale invariance and which exhibit well-defined thresholds given by the  $(N-1)$-body subsystem.} with $N>3$ exist, because of the quite different scaling and threshold properties of the cluster states \cite{Amado1973tin}. However, other approaches to extend universal Efimov theory to larger systems have been pursued along different lines \cite{Sorensen2002ctb, Yamashita2006fbs, Thogersen2008nbe}. The development of accurate descriptions of four-boson systems \cite{Hammer2007upo,Vonstecher2009sou,Deltuva2010epi} demonstrates the existence of four-body states tied to each three-body Efimov state. In a major extension of Efimov physics, Ref.\,\cite{Vonstecher2010wbc} predicts the existence of a family of cluster states tied to an Efimov trimer.

According to Refs.\,\cite{Vonstecher2009sou, Deltuva2010epi, Vonstecher2010wbc, Vonstecher2011fas}, the binding energy of the cluster states follows universal scaling laws, which are directly connected to the Efimov effect. Figure\,\ref{Fig.1} shows the calculated energy spectrum for the ground states of three-, four- and five-body clusters (lower panel). The corresponding experimental observables are loss peaks in the atom number (upper panel), which appear at values of the scattering length $a_-$, $a_{4,-}$, and $a_{5,-}$, where the three-, four- and five-body states cross the free atom threshold, respectively. The resonant values of the scattering length are predicted to be universally connected by the simple relations $a_{4,-}=0.44(1)\,a_-$ and $a_{5,-}=0.65(1)\,a_{4,-}$ \cite{Vonstecher2011fas}. A more recent study \cite{Deltuva2012ubt} has theoretically explored the four-boson resonance and has determined its position with greater accuracy. For four-body states, the universal relation has been confirmed in experiments \cite{Zaccanti2009ooa, Pollack2009uit, Ferlaino2009efu, Ferlaino2011eri} and the four-body recombination rate has been measured \cite{Ferlaino2009efu} and calculated within the hyperspherical framework \cite{Vonstecher2009sou}. A straightforward extension to five-body systems is not currently possible for experiment nor theory. The experimental challenge is to discriminate the five-body recombination signal against a strong background resulting from fewer-body processes. The numerical difficulty of the scattering few-body problem grows exponentially with the number of particles making the description of five and larger systems beyond current theoretical capabilities.

This article presents a combined theoretical and experimental study of universal few-body physics up to five-body states. We present strong evidence for the existence of an Efimov-related cluster state of five identical bosons and we provide quantitative results for the corresponding five-body recombination rate. Our results highlight a new level of understanding concerning few-body physics and its experimental manifestations in ultracold atomic quantum gases.

\section{Theoretical approach}

The theoretical analysis of $N$-body recombination processes requires the description of the $N$-body scattering continuum. The hyperspherical framework has been successfully applied to describe recombination processes for $N>3$ \cite{Vonstecher2009sou,Mehta2009gtd}. In this framework, the Hamiltonian is diagonalized adiabatically as a function of the hyperradius $R$, which describes the overall size of the system, leading to a set of coupled one-dimensional Schr\"odinger equations. At ultralow temperatures and large scattering lengths, $N$-body recombination events are mainly controlled by scattering processes with incoming flux in the lowest $N$-body scattering channel and outgoing flux in deeper loss channels. The coupling to the deep channels is assumed to remain approximately unaffected as the scattering length and the collision energy are varied throughout our regime of interest. Thus, all the relevant information comes from the analysis of the lowest $N$-body potential curve corresponding to the incoming scattering channel.

However, the extraction of the hyperspherical potential curves and couplings becomes computationally unfeasible as the number of particles increases. Currently, the numerical technology allows for the calculation of potential curves of three and four-body systems but no tractable method has yet been implemented that can calculate both the potential curves and couplings for $N>4$. However, the trapped energy eigenvalue spectrum of a five-body system can be accurately obtained with current technology. Thus, the present study extracts the relevant recombination information from an analysis of the trapped spectrum.

Our starting point is the hyperspherical description of the ultracold $N$-body recombination rate \cite{Mehta2009gtd},
\begin{equation}
L_N^{0^+}=\left(\frac{4\pi}{k^2}\right)^{(3 N-5)/2} \frac{\hbar N \Gamma(3N/2-3/2) }{\mu_N} (1-|S_{00}^{0^+}|^2),
\label{ln}
\end{equation}
where $\mu_N=m/\sqrt[N-1]{N}$ with $m$ being the atomic mass, $k=(2\mu_N E/\hbar^2)^{1/2}$ is the incoming hyperradial scattering wavenumber and $S_{00}^{0^+}$ is the diagonal element of the S-matrix for the lowest channel (00) in the $J^\Pi=0^+$ symmetry. For purely elastic scattering $|S_{00}^{0^+}|^2=1$ and the recombination rate is zero. In the limit in which every $N$-body collision leads to losses, $|S_{00}^{0^+}|^2=0$. Taking the thermal average in the full loss case at a temperature $T$, one obtains the unitary limit for $N$-body recombination at low energy:
\begin{equation}
\langle L_N \rangle_T=(2\pi)^{(3 N-5)/2} N \frac{k_B T}{\hbar} \left(\frac{\hbar^2 }{\mu_N k_B T}\right)^{3( N-1)/2}.
\label{Uni}
\end{equation}

\begin{figure}
\begin{center}
\includegraphics[width=.6\textwidth]{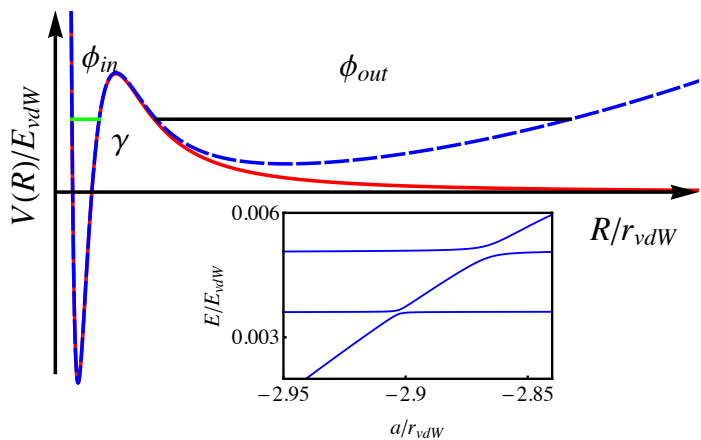}
\caption{Schematic representation of lowest five-body hyperspherical potential curve. The solid curve represents the free-space hyperspherical potential while dashed curves represent hyperspherical potential curves in the presence of a harmonic potential. Solid lines represent the inner and outer WKB phases. Inset: Five-body trapped energy spectrum in the region where the bound five-body state crosses the lowest trapped states.}
\label{FigNew}
\end{center}
\end{figure}

Based on potential curves computed for the three- and four-body cases, we expect that the five-body potential curve should have the topology depicted as a solid curve in Fig.\,\ref{FigNew}, i.e., the lowest potential curve exhibits a barrier that separates the inner region (small $R$) from the asymptotic scattering region at large $R$. Note that the effective mock-centrifugal barrier \cite{Fano1976doe} in the lowest $N$-body continuum channel at large $R$ is guaranteed to have the form (for finite $a$), $U(R) \rightarrow \hbar^2 (3N-4)(3N-6)/2 \mu_N R^2$.
For this potential curve topology, an extension of the semiclassical (WKBJ) treatment of Berry \cite{Berry1966scs} to include the decay to the lowest channels yields \cite{Mehta2009gtd}:
\begin{equation}
(1-|S_{00}|^2)=\frac{e^{-2\gamma}}{2} \frac{\sinh(2\eta_-)}{\cos^2\phi_{\rm in}+\sinh^2\eta_-}A(\eta_-,\gamma,\phi_{\rm in}),
\label{1s}
\end{equation}
where $\phi_{\rm in}$ is the WKBJ phase for the inner allowed region, $\gamma$ is the WKBJ tunneling integral in the barrier region, $\eta_-$ describes the decay to deeper, non-universal, channels and is treated as a fitting parameter, and $A(\eta_-,\gamma,\phi_{\rm in})$ ensures the proper normalization.
 Thus, the determination of $\gamma$ and $\phi_{\rm in}$ for the relevant range of energy and scattering length gives an approximate description of five-body recombination.

The novelty of our approach is the determination of $\gamma$ and $\phi_{\rm in}$ from an analysis of the trapped spectrum.
The five-body trapped energy spectrum is obtained using a correlated Gaussian basis set expansion \cite{Vonstecher2011fas}; for details see \ref{app1}. In the region where the five-body resonance occurs, the spectrum exhibits a series of avoided crossings between the five-body states bound in the inner region and the outer trap region states (see Fig.\,\ref{FigNew}).
In the WKBJ approximation, the quantization condition for the trapped potential curve (dashed curve in Fig.\ref{FigNew}) is $\frac{\Delta}{2}\tan (\phi_{\rm in} )\tan (\phi_{\rm out} )=1$, where $\Delta=e^{-2 \gamma }/2$.
For collision energies below the barrier local maximum and away from the avoided crossings, the allowed energy eigenvalues occur when $\phi_\alpha (E,a)\approx\pi(i+1/2)$, where $i$ is an integer and $\alpha=\mbox{in, out}$.
The phase $\phi_{\rm in}$ in the four-boson case near the resonance energy is known from our previous work~\cite{Mehta2009gtd} to be well-described by $\phi_{\rm in}\approx \phi_{\rm in,0} + b (a/r_{\rm vdW})+c a E/(r_{\rm vdW}E_{\rm vdW})$, where $\phi_{\rm in,0}$, $b$ and $c$ are fitting parameters and $r_{\rm vdW}$ and $E_{\rm vdW}$ are the van der Waals radius and energy, respectively, as defined in \cite{Chin2010fri}. Using this simple form in the five-body case and imposing the eigenstate condition for bound states, the values of $\phi_{\rm in,0}$, $b$ and $c$ for $N=5$ are extracted.

Next, an analysis of the spectrum at the avoided crossings (see e.g.\ Fig.\,\ref{FigNew}) determines $\gamma$, as explained in detail in \ref{app2}. The relevant avoided crossings occur when $\phi_{\rm in}\approx \phi_{\rm out}\approx \pi(i+1/2)$. For narrow avoided crossings ($\Delta\ll1$), the quantization condition reduces to $\delta\phi_{\rm in}\, \delta\phi_{\rm out} \approx\frac{\Delta}{2}$ where $\phi_\alpha=\pi(i+1/2)+\delta\phi_\alpha$. Right at the avoided crossing, the energy difference between the two states (2$\Delta E$) is related to $\Delta$, namely as $\delta\phi_\alpha\approx (d\phi_\alpha/dE ) \Delta E$. The quantization condition thus reduces to $\Delta\approx 2 \Delta E^2 (d\phi_{\rm in}/dE )(d\phi_{\rm out}/dE )$. Consequently knowledge of $\phi_{\rm in}$, $\phi_{\rm out}$ and the energy avoided crossings allows the tunneling $\gamma$ to be determined. At large $R$, interactions can be treated perturbatively leading to an hyperspherical potential curve valid at large $R$ that determines $\phi_{\rm out}$.

By changing the trapping confinement we can change $\phi_{\rm out}$ and by introducing short range three-body forces we can modify $\phi_{\rm in}$ without affecting the barrier. This allow us to explore how $\gamma$ depends on both $E$ and $a$. Interestingly, at low energies our numerical results are nicely fitted by the formula: $e^{-2\gamma}\propto (E/E_{\rm vdW})^5|a/r_{\rm vdW}|^{9.6}$, which is in good agreement with the predicted threshold behavior $(1-|S_{00}^{0^+}|^2)\propto E^5 |a|^{10}$ \cite{Mehta2009gtd}.

\section{Experiment}

We prepare an optically trapped sample of cesium atoms in the lowest sub-level of the electronic ground state under similar conditions as described in Ref.\,\cite{Berninger2011uot}. The final evaporation process, performed at a magnetic field of 894\,G ($a = +285\,a_0$, where $a_0$ is the Bohr radius), is stopped before the onset of Bose-Einstein condensation. The sample is then adiabatically recompressed to avoid further evaporation from the trap. At this point, the sample contains about $6\times10^4$ atoms at a temperature $T = 78(3)\,\rm{nK}$ and the confining optical potential has a mean trap frequency $\bar{\omega} = 2\pi \times36.2(2)$\,Hz, which results in a peak number density of $4.2\,\mu\rm{m}^{-3}$ and a peak phase space density close to unity. The wide $s$-wave Feshbach resonance with its pole at 786\,G offers ideal tuning properties \cite{Berninger2011uot, Berninger2012frw}, superior to the low-field region investigated in our previous works \cite{Kraemer2006efe, Ferlaino2009efu}.

We measured the decay of the atom number in the region of negative scattering length (from $-500\,a_0$ at 863\,G to $-200\,a_0$ at 873\,G), where the four- and five-body recombination resonances are expected. After the recompression stage, we tune the scattering length to its target value, and we measure the atom number after a variable hold time by absorption imaging. After 100\,ms, the typical loss fraction is around 10\% at $-300\,a_0$ and almost 35\% at $-450\,a_0$.

The time evolution of the number $\mathcal{N}$ of trapped atoms and temperature $T$ are determined by the different $N$-body loss processes and can be expressed \cite{Weber2003tbr} in terms of a system of coupled differential equations,
\begin{equation}
\dot{\mathcal{N}}/\mathcal{N}=-\sum^{+\infty}_{N=1}L_N\langle n^{N-1} \rangle , \label{diff}
\end{equation}
\begin{equation}
\dot{T}/T=\sum^{+\infty}_{N=1} \epsilon_N L_N \langle n^{N-1} \rangle ,
\label{diffT}
\end{equation}
where $L_N$ represents the $N$-body rate coefficients. The averaged atom densities are evaluated as $\langle n^{N-1}\rangle=\int n^N d^3\bf{r}$ $= \mathcal{N}^{N-1} N^{-3/2} [(m \bar{\omega}^2)/(2\pi k_BT)]^{(3N-3)/2}$ from a thermal distribution of the harmonically trapped atoms. Equation (\ref{diffT}) incorporates the anti-evaporation heating \cite{Weber2003tbr} that results from the higher recombination rate in the densest part of the cloud ($2\epsilon_N\equiv1-1/N$).

Under our experimental conditions, the first two terms of the sums can be omitted. One-body losses, resulting from collisions with the background gas, are negligible on our experimental time scale ($L_1=0$), and the use of atoms in their lowest sub-state assures that two-body losses vanish ($L_2=0$). Therefore, the three-body rate coefficient $L_3$ is the leading term contributing to the atom losses, and its resonant behavior is related to the Efimov trimers.
This contribution is well understood and the coefficient $L_3$ can be described by the well-established result of effective field theory \cite{Braaten2001tbr}. The rates of recombination events involving more particles are generally smaller than the one related to three-body losses and the contributions are difficult to discriminate because of the very similar behavior. Since the rate of recombination events for typical gas densities decreases rapidly with $N$, contributions with $N>5$ are considered negligible in the following.

\begin{figure}
\begin{center}
\includegraphics[width=.65\textwidth] {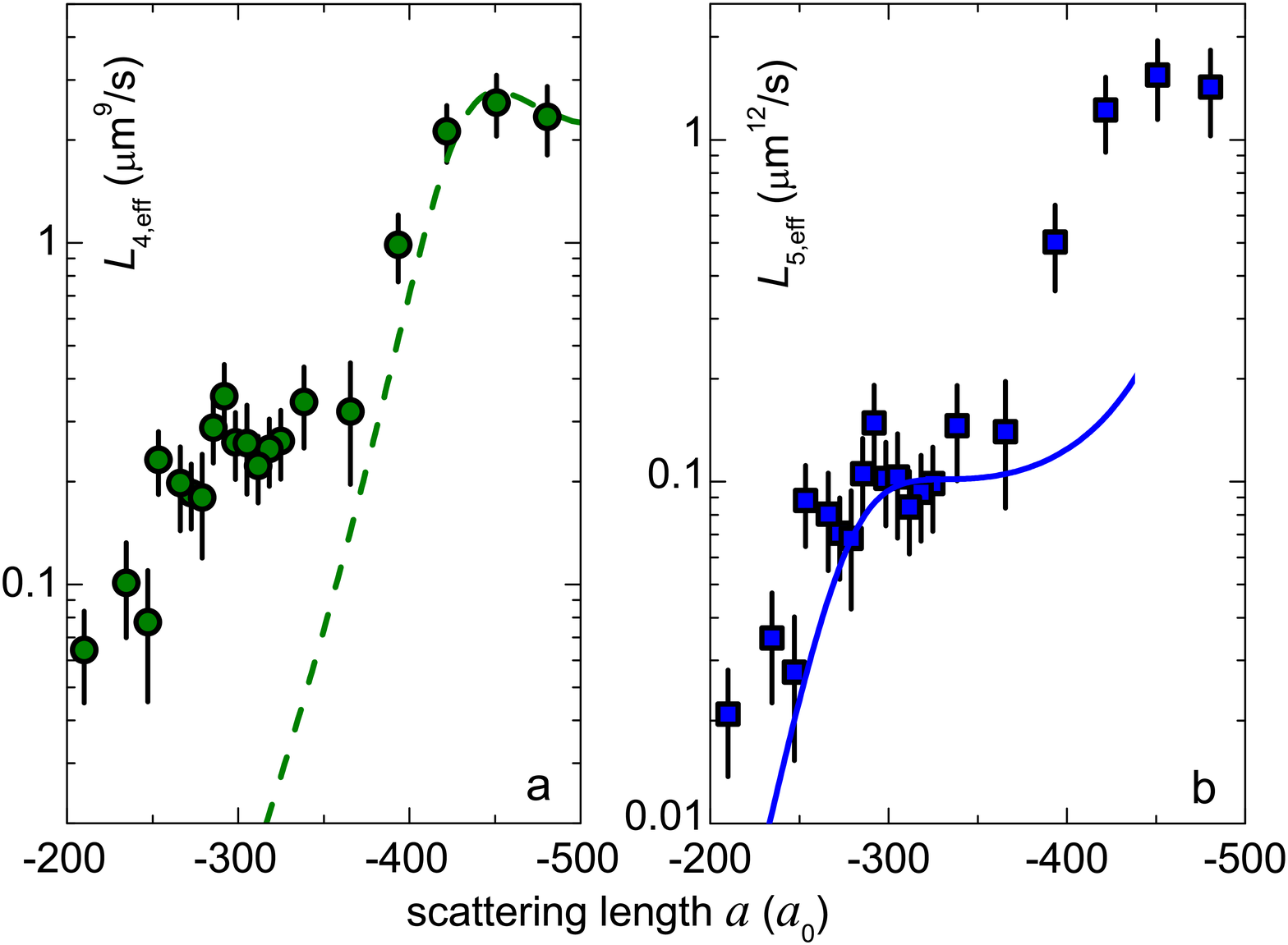}
\caption{Effective four- (a) and five-body recombination rates (b). The green dashed curve and the blue solid line follow the theoretical model for $L_4$ and $L_5$, respectively, with an additional scaling factor for $L_5$; see text. The error bars include the statistical uncertainties from the fitting routine, the temperature and the trap frequencies.}
\label{Fig.3}
\end{center}
\end{figure}

A general fit to the experimental decay curves with $L_3$, $L_4$, and $L_5$ as free parameters in Eq.\,(\ref{diff}) and Eq.\,(\ref{diffT}) turns out to be practically impossible. Therefore, we fix $L_3$ according to effective field theory, with parameters $a_-=-955\,a_0$ and $\eta_-=0.08$ as determined for three-body recombination ($N=3$) in our previous experiment \cite{Berninger2011uot}. We can now interpret the additional losses in terms of four-body and five-body decay. In order to avoid any fitting ambiguities, we chose the simple approach to describe these losses either in terms of an effective four-body loss coefficient $L_{4,\rm{eff}}$ (setting $L_5 = 0$) or an effective five-body loss coefficient $L_{5,\rm{eff}}$ (setting $L_4 = 0$).

Figure \ref{Fig.3}(a) shows $L_{4,\rm{eff}}$ as extracted from our experimental data in comparison with the theoretical predictions, obtained by numerically evaluating $L_4$ from Eqs.\,(\ref{ln}) and (\ref{1s}) for our experimental conditions. Here we have adjusted the decay parameter to $\eta_-=0.33$, which as a non-universal parameter depends on $N$. The comparison shows that the losses observed around $-450\,a_0$ can be fully attributed to the four-body recombination resonance. The four-body loss peak position $a_{4,-}=-440(10)\,a_0$ corresponds to $0.46(1)\,a_-$ and is in very good agreement with the theoretical value $0.44(1)\,a_-$ \cite{Vonstecher2009sou} and previous observations \cite{Ferlaino2009efu, Ferlaino2011eri, Pollack2009uit, Zaccanti2009ooa}. In contrast the enhancement of losses centered at about $-300\,a_0$ cannot be interpreted in terms of the known universal four-body cluster states\footnote{We cannot rule out the possibility of a non-universal few-body state, which may be associated with higher partial waves. However, such an accidental coincidence appears to be rather unlikely.}, suggesting that a different loss mechanism is present.

An alternative representation of the same data in terms of $L_{5,\rm{eff}}$ is shown in Fig.\,\ref{Fig.3}(b), together with the results of our theoretical model for $L_5$; here we adjusted the relevant decay parameter to $\eta_-=0.20$. The model nicely explains the loss rates in the region where three- and four-body losses cannot account for the experimental observations. Remarkably, the resonance position $a_{5,-}\,=0.64(2)\,a_{4,-}$ is in agreement with the theoretical predictions $0.65(1)\,a_{4,-}$ \cite{Vonstecher2010wbc, Vonstecher2011fas}. However, quantitatively, the experimental values for $L_5$ are about 15 times larger than the calculated ones. To account for this, we introduce a corresponding scaling factor.
We find that this deviation might derive from non-universal effects that modify the value of the calculated WKB phase $\gamma$ by about 10\%, which remains within the realistic uncertainty range of our theory.

An experimental search for higher-order recombination resonances ($N>5$) that would be expected at lower values of the scattering length did not show clear signatures. A general problem arises, namely that the phase-space density cannot be further increased without causing the collapse of a Bose-Einstein condensate at negative values of $a$. To induce faster losses, adiabatic compression can increase the density, but then the higher temperatures cause increasing problems with the unitarity limit for high $N$. By decreasing the temperature, constraints by the unitarity limit can be avoided, but then losses for high $N$ get so small that they become practically unobservable.

\begin{figure}
\begin{center}
\includegraphics[width=.6\textwidth]{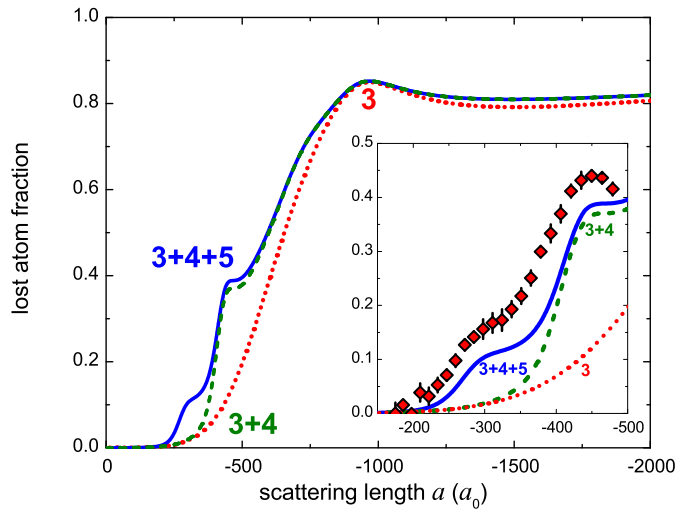}
\caption{Calculated and measured fraction of loss atoms from an atomic sample of initially $5\,\times\,10^4$ atoms at a temperature of 80\,nK after a hold time of 100\,ms. The red dotted line corresponds to the losses predicted for three-body recombination only, while the dashed green line and the blue solid line include also contributions from four- and five body recombination, as quantified in this work. A cut-off to the maximum losses has been applied according to Eq.\,(\ref{Uni}), as suggested in \cite{Greene2004arf}. The inset shows a close up of the region of dominant four- and five-body recombination and compares the theoretical results with the experimental observations.}
\label{Fig.2}
\end{center}
\end{figure}

Based on the above results and parameters, we model the general loss behavior in the region of interest. Figure \ref{Fig.2} shows an example for three, four- and five-body recombination by plotting the atom losses, under typical experimental conditions, for a fixed hold time and variable scattering length. The ``family portrait" of $N$-body recombination highlights the different contributions and confirms how the different loss features dominate the losses at the resonant positions. The experimental data plotted in the inset show that the peak positions and the magnitude of losses are in very good agreement with the simulated losses. Note that the somewhat higher experimental losses can be attributed to an additional loss that occurs during the ramp to the target magnetic field strength.

\section{Conclusion}

By pushing the limits of ultracold few-body physics, we have explored a universal five-body recombination resonance both experimentally and theoretically. The observed series of recombination features, which we interpret in terms of three-, four- and five-body recombination resonances, provides crucial evidence for the existence of a family of universal $N$-body bound states tied to Efimov trimers. The infinite series of $N$-boson cluster states represents a paradigm for the general implications of Efimov physics for many-body systems. We speculate that similar scenarios also exist for other few-body systems of increasing size, containing fermionic constituents or particles of different masses, with important consequences for the interaction properties of the many-body system.

\section*{Acknowledgments}

This work was supported by the Austrian Science Fund FWF within project P23106, and in part by the U.S. National Science Foundation. A. Z. was supported within the Marie Curie Project LatTriCs 254987 of the European Council.

\appendix

\section{Extraction of the trapped five-body spectrum}
\label{app1}

A crucial ingredient of our recombination analysis is the accurate extraction of the five-body trapped spectrum. The trapped few-boson Hamiltonian is given by
\begin{equation}
\mathcal{H}=\sum_{i=1}^N \left( -\frac{\hbar^2\nabla^2_i}{2m}+\frac{m\omega^2r_i^2 }{2}\right)+\sum_{i>j} V_{2b} (r_{ij})+\sum_{i>j>k} V_{3b} (R_{ijk})
\end{equation}
where $\omega$ is the trapping frequency, $V_{2b}(r)=V_{2b0} e^{-r^2/(2r_0^2)}$ is the two-body potential and $V_{3b}(R)=V_{3b0} e^{-R^2/(2R_0^2)}$ is a three-body potential. Here, $r_0$ and $R_0$ are the ranges of the two and three-body potentials that are fixed during the calculation, and $R_{ijk}=\sqrt{(r_{ij}^2+r_{ik}^2+r_{jk}^2)/3}$.  The two-body potential strength $V_{2b0}$ is used to tune the two-body scattering length and $V_{3b}$ is set to zero in most calculations (see discussion below). To relate the model potential with the experimental we rely on the universal character of the low energy few-boson physics and we relate $r_0$ with the van der Waals length  $r_{vdw}$ so that the three- and four-body recombination peaks coincide with those experimentally observed. In this transformation, we relate the energy scales so that they scale as inverse length squared.

To extract the recombination parameters, the five-body spectrum is analyzed as a function of the scattering length and the trapping confinement.  The scattering length is changed in the region $-4 r_0< a<0$ and the trapping frequency is changed so that the corresponding trapping length $a_{ho}=\sqrt{\hbar/(m\omega)}$ varies in the range  $5 r_0<a_{ho} \le 100 r_0$. This region of scattering lengths and energies corresponds to the low energy region where the five-body resonance occurs.

The three-body interaction is used to shift the position of the five-body resonance and explore the dependence of the recombination on the scattering length. The range of the three-body potential is taken to be $R_0$, which is smaller than the two-body $r_0$ so that it is mainly relevant at small hyperradii, i.e. so that its main contribution in our recombination formula is to change the inner phase.
For the three-body interactions considered, the spectrum follows the linear dependence with the scattering length that is expected at small and negative scattering lengths. This linear dependence arises from only two-body physics which, in the hyperspherical framework, is described by the long-range behavior of the potential curves.  Using the zero-range model of the two-body interaction, one can derive a first order correction of the hyperspherical potential curve. Here, we follow a similar procedure to that on Ref.~\cite{Bohn1998epf}, but for a set of coordinates in which the center of mass has been removed. In this approximation, the lowest potential takes the form
\begin{equation}
\fl V(R)=\frac{3 (N-2) (3 N-4)\hbar^2}{8 \mu_N R^2}+\frac{\mu_N \omega^2R^2}{2} +\frac{\hbar^2 a (N-1) N^{\frac{2N-1}{2N-2}} \Gamma \left(\frac{3}{2} N - \frac{3}{2} \right)}{\sqrt{2 \pi } \mu_N R^3 \Gamma
   \left(\frac{3}{2} N - 3\right)}.
\label{HRpot}
\end{equation}
The first two terms correspond, respectively, to the hyperangular or ``mock-centrifugal'' kinetic energy and the trapping potential; and the third term represents the interaction corrections.
Equation~\ref{HRpot} is {\em only} valid for small $|a|$ in the region where $R$ is much larger than the interaction range.  Using perturbation analysis, one derives the well-known corrected energy of the lowest state,
\begin{equation}
E\approx 3(N-1)\frac{\hbar\omega}{2}+\sqrt{\frac{2}{\pi}} \frac{a}{a_{ho}}  \hbar\omega \frac{N(N-1)}{2}.
\label{Elinear}
\end{equation}
Our numerical calculations, with and without the three-body forces, show the linear behavior described in Eq.~\ref{Elinear} in the region $|a|<|a_{5,-}|$. This suggests that the hyperradial potential in the $R\gg r_0$ is well described by Eq.~\ref{HRpot} and that, in this region, the potential is independent of the three-body forces.
 Thus,  it is consistent to interpret the main contribution of the three-body interaction as a modification of the short range physics that controls $\phi_{in}$.

\begin{figure}
\begin{center}
\begin{tabular}{ccc}
\includegraphics[width=\textwidth]{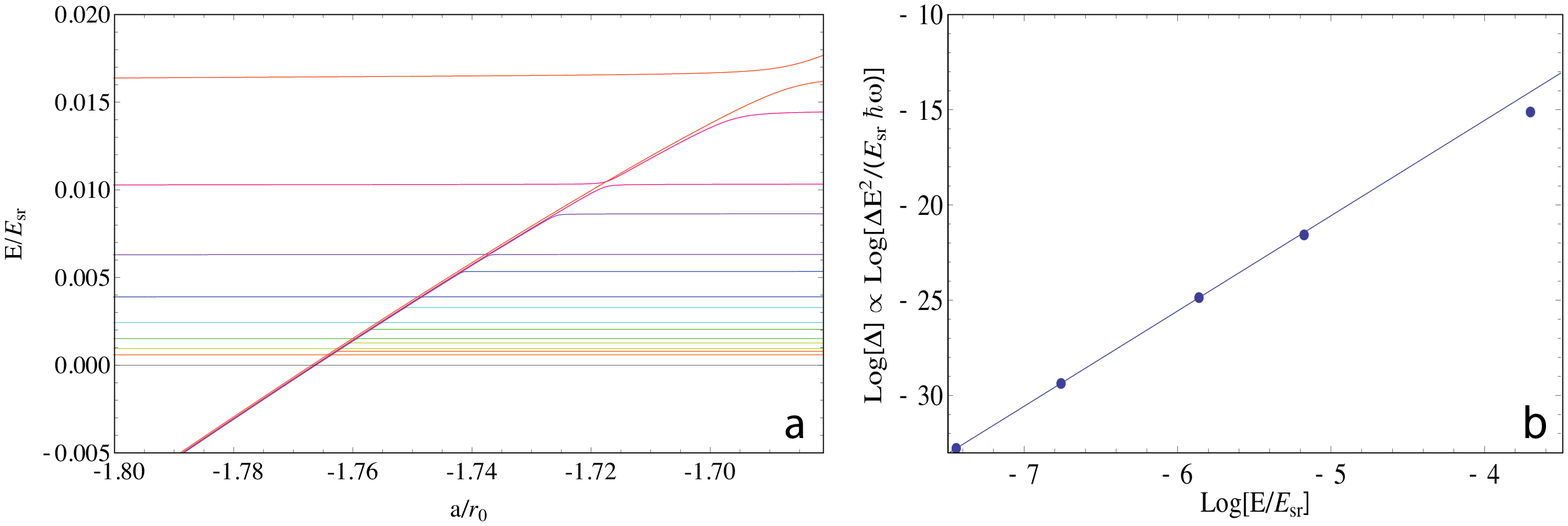}
\end{tabular}
\caption{(a) The lowest two eigenenergies of a trapped five body system are shown as functions of the scattering length for different trapping frequencies. Different colors represent different trapping frequencies. The combination of these states essentially describes the energy of the five-body state in the inner region of the potential $E_{mol}(a)$ (the diagonal curve). Here $E_{sr}=\hbar^2/(m r_0^2)$ and $r_0$ is the characteristic range of the two-body model potential that can be tuned to obtain the five-body resonance (i.e. $r_0\sim 1.7 r_{vdw}$ where $r_{vdw}$ is the van der Waals length). (b) The near-threshold behavior of $\Delta$. The fitting of the lowest energy points leads that $\Delta\propto A E^b$.  The lowest three points lead to $b\approx 5.004$ as expected from the known threshold behavior \cite{Mehta2009gtd}.}
\label{S1}
\end{center}
\end{figure}

The calculations are carried out using a correlated Gaussian basis set expansion limited to describe $L^P=0^+$ trapped states since the energetically lowest scattering continuum corresponds to zero angular momentum and positive parity. In this computation, we are only interested in the lowest trapped states which, in the hyperspherical picture, are supported mainly by the lowest potential curve. Therefore, the basis set is designed to accurately describe those states. The calculations include thousands of basis functions which are optimized for different scattering lengths and trap lengths. To verify the convergence of the energies, we carry out several optimization steps. The typical spectrum obtained by this analysis is shown in Fig.\,\ref{S1}(a).

\section{Semiclassical analysis of the trapped spectrum}
\label{app2}

To extract the semiclassical recombination parameters, we repeat the same semiclassical analysis used to derive Eqs.~(2,3) but for a hyperspherical potential with the trapping potential (see Fig.~\ref{FigNew}). This basically amounts to solve a double well problem using a semiclassical analysis. For this analysis, we follow closely the prescription in Ref.~\cite{Berry1966scs} to determine the quantization condition
\begin{equation}
\beta=-\tan ^{-1}\left(\frac{1}{4} e^{-2 \gamma } \tan (\phi_{in} )\right)-\phi _{out}+\frac{\pi}{2}=n\pi
\end{equation}
where $\phi_{in}$ is the phase in the inner region, $\gamma$ is the barrier phase and $\phi_{out}$ is the phase in the external trapped region (see Fig.~\ref{FigNew}). The semiclassical phases  are
\begin{eqnarray}
\phi_{in}= \int_{\mbox{inner well}} q(R) dR, \\
\gamma(E,a)=\mbox{Im} \int_{\mbox{barrier}} q(R) dR, \\
\phi_{out}= \int_{\mbox{trap well}} q(R) dR,
\end{eqnarray}
where $q(R)=\sqrt{2\mu[E-V(R)]}$ and $V(R)$ is the hyperspherical curve with the Langer correction. Here, the different integration regions are bounded by the classical turning points, i.e. the $R$ positions at which $q(R)=0$.
In our analysis, $\phi$ and $\gamma$ are assumed to be unaffected by the trapping potential which is expected to be negligible at small hyperradii.

After some mathematical manipulation, the quantization condition can be written as
\begin{equation}
\frac{1}{\tan (\phi_{in} )\tan (\phi_{out} )}=\frac{\Delta}{2}
\label{QuantCond}
\end{equation}
where $\Delta=e^{-2 \gamma }/2$. For tunneling energies well below the barrier height $\Delta\ll1$ which implies that the quantization condition is fulfilled when either $\phi_{in}\approx\pi(i_{in}+1/2)$ or  $\phi_{out}\approx\pi(i_{out}+1/2)$. These conditions ($\phi_{in}\approx\pi(i_{in}+1/2)$ and  $\phi_{out}\approx\pi(i_{out}+1/2)$) can be interpreted as having a bound state either in the inner or outer region. We call $E_{in}(a)$ and $E_{out}(a)$ the energies at which $\phi_{in}=\pi(i_{in}+1/2)$ and  $\phi_{out}=\pi(i_{out}+1/2)$, respectively. Here $i_\alpha$ are integers representing the number of bound states supported by the inner and outer well respectively. Thus, close to those energies, the inner and outer phases take the form $\phi_{\alpha}= \pi(i_{\alpha}+1/2)+\phi_\alpha' (E-E_{\alpha}(a))$ where $\phi_\alpha' \equiv d\phi_\alpha /dE$. In the vicinity of the avoided crossings, the quantization condition can be approximated to
\begin{equation}
(E_{in}(a)-E)(E_{out}(a)-E)-\frac{\Delta}{2 \phi_{in}' \phi_{out}'}\approx 0.
\label{QuantCond2}
\end{equation}

Equation~\ref{QuantCond2} resembles a determinant of a $2\times2$ Hamiltonian matrix which lends itself to the following physical interpretation. The states supported by the inner and outer regions are coupled to each other and the coupling energy $V_{in,out}$ is related to the tunneling through the barrier by $V_{in,out}^2=\frac{\Delta}{2 \phi_{in}' \phi_{out}'}$.  The eigenenergies of this double well problem ($E_1$ and $E_2$) reach their squared minimal difference at the avoided crossing, equal to:
\begin{equation}
(E_2-E_1)^2=2\Delta/(\phi_{in}' \phi_{out}').
\end{equation}

To verify the validity of this equation, the energy dependence of $\Delta$ is explored by varying the strength of the trapping potential. According to the threshold laws \cite{Mehta2009gtd}, for a five-boson problem, $\Delta$ increases with energy as $\Delta\propto E^5$.  Figure \ref{S1}(b) confirms that the scaling of $\Delta$ calculated through the avoided crossing analysis shows excellent agreement with the threshold law prediction.

\section*{References}

\bibliographystyle{iopart-num}

\begin{thebibliography}{}
\expandafter\ifx\csname url\endcsname\relax
  \def\url#1{{\tt #1}}\fi
\expandafter\ifx\csname urlprefix\endcsname\relax\def\urlprefix{URL }\fi
\providecommand{\eprint}[2][]{\url{#2}}

\bibitem{Braaten2006uif}
Braaten E and Hammer H~W 2006 {\em Phys. Rep.\/} {\bf 428} 259--390

\bibitem{Ferlaino2010fyo}
Ferlaino F and Grimm R 2010 {\em Physics\/} {\bf 3} 9

\bibitem{greene2010uif}
Greene C~H 2010 {\em Phys. Today\/} {\bf 63(3)} 40--45

\bibitem{Chin2010fri}
Chin C, Grimm R, Julienne P~S and Tiesinga E 2010 {\em Rev. Mod. Phys.\/} {\bf
  82} 1225--1286

\bibitem{Efimov1970ela}
Efimov V 1970 {\em Phys. Lett. B\/} {\bf 33} 563--564

\bibitem{Nielsen1999ler}
Nielsen E and Macek J~H 1999 {\em Phys. Rev. Lett.\/} {\bf 83} 1751--1754

\bibitem{Esry1999rot}
Esry B~D, Greene C~H and Burke J~P 1999 {\em Phys. Rev. Lett.\/} {\bf 83}
  1751--1754

\bibitem{Kraemer2006efe}
Kraemer T, Mark M, Waldburger P, Danzl J~G, Chin C, Engeser B, Lange A~D, Pilch
  K, Jaakkola A, N\"agerl H~C and Grimm R 2006 {\em Nature\/} {\bf 440}
  315--318

\bibitem{Ottenstein2008cso}
Ottenstein T~B, Lompe T, Kohnen M, Wenz A~N and Jochim S 2008 {\em Phys. Rev.
  Lett.\/} {\bf 101} 203202

\bibitem{Huckans2009tbr}
Huckans J~H, Williams J~R, Hazlett E~L, Stites R~W and O'Hara K~M 2009 {\em
  Phys. Rev. Lett.\/} {\bf 102} 165302

\bibitem{Knoop2009ooa}
Knoop S, Ferlaino F, Mark M, Berninger M, Sch\"obel H, N\"agerl H~C and Grimm R
  2009 {\em Nature Phys.\/} {\bf 5} 227--230

\bibitem{Zaccanti2009ooa}
Zaccanti M, Deissler B, D'Errico C, Fattori M, Jona-Lasinio M, M�ller S, Roati
  G, Inguscio M and Modugno G 2009 {\em Nature Phys.\/} {\bf 5} 586

\bibitem{Barontini2009ooh}
Barontini G, Weber C, Rabatti F, Catani J, Thalhammer G, Inguscio M and Minardi
  F 2009 {\em Phys. Rev. Lett.\/} {\bf 103} 043201

\bibitem{Gross2009oou}
Gross N, Shotan Z, Kokkelmans S and Khaykovich L 2009 {\em Phys. Rev. Lett.\/}
  {\bf 103} 163202

\bibitem{Nakajima2010nea}
Nakajima S, Horikoshi M, Mukaiyama T, Naidon P and Ueda M 2010 {\em Phys. Rev.
  Lett\/} {\bf 105} 023201

\bibitem{Gross2010nsi}
Gross N, Shotan Z, Kokkelmans S and Khaykovich L 2010 {\em Phys. Rev. Lett\/}
  {\bf 105} 103203

\bibitem{Lompe2010ads}
Lompe T, Ottenstein T~B, Serwane F, Viering K, Wenz A~N, Z\"{u}rn G and Jochim
  S 2010 {\em Phys. Rev. Lett\/} {\bf 105} 103201

\bibitem{Nakajima2011moa}
Nakajima S, Horikoshi M, Naidon T~M~P and Ueda M 2011 {\em Phys. Rev. Lett.\/}
  {\bf 106} 143201

\bibitem{Pollack2009uit}
Pollack S~E, Dries D and Hulet R~G 2009 {\em Science\/} {\bf 326} 1683--1686

\bibitem{Ferlaino2011eri}
Ferlaino F, Zenesini A, Berninger M, Huang B, N\"{a}gerl H~C and Grimm R 2011
  {\em Few-Body Systems\/} {\bf 51} 113--133

\bibitem{Blume2000mch}
Blume D and Greene C~H 2000 {\em J. Chem. Phys.\/} {\bf 112} 8053--8067

\bibitem{Blume2002foa}
Blume D, Esry B~D, Greene C~H, Klausen N~N and Hanna G~J 2002 {\em Phys. Rev. Lett.\/} {\bf
  89} 163402

\bibitem{Adhikari1981fbe}
Adhikari S~K and Fonseca A~C 1981 {\em Phys. Rev. D\/} {\bf 24} 416--425

\bibitem{Naus1987tee}
Naus H~W~L and Tjon J~A 1987 {\em Few-Body Syst.\/} {\bf 2} 121--126

\bibitem{Platter2004fbs}
Platter L, Hammer H~W and Mei\ss{}ner U~G 2004 {\em Phys. Rev. A\/} {\bf 70}
  052101

\bibitem{Hanna2006eas}
Hanna G~J and Blume D 2006 {\em Phys. Rev. A\/} {\bf 74} 063604

\bibitem{Richard1994lot}
Richard J~M and Fleck S 1994 {\em Phys. Rev. Lett.\/} {\bf 73} 1464--1467

\bibitem{Yamashita2011bae}
Yamashita M~T, Fedorov D~V and Jensen A~S 2011 {\em Few-Body Syst.\/} {\bf 51}
  135--151

\bibitem{Amado1973tin}
Amado R~D and Greenwood F~C 1973 {\em Phys. Rev. D\/} {\bf 7} 2517

\bibitem{Sorensen2002ctb}
S\o{}rensen O, Fedorov D~V and Jensen A~S 2002 {\em Phys. Rev. Lett.\/} {\bf
  89} 173002

\bibitem{Yamashita2006fbs}
Yamashita M~T, Tomio L, Delfino A and Frederico T 2006 {\em Europhys. Lett.\/}
  {\bf 75} 555--561

\bibitem{Thogersen2008nbe}
Th{\o}gersen M, Fedorov D~V and Jensen A~S 2008 {\em Europhys. Lett.\/} {\bf
  83} 30012

\bibitem{Hammer2007upo}
Hammer H~W and Platter L 2007 {\em Eur. Phys. J. A\/} {\bf 32} 113--120

\bibitem{Vonstecher2009sou}
{von Stecher} J, D'Incao J~P and Greene C~H 2009 {\em Nature Phys.\/} {\bf 5}
  417--421

\bibitem{Deltuva2010epi}
Deltuva A 2010 {\em Phys. Rev. A\/} {\bf 82} 040701

\bibitem{Vonstecher2010wbc}
{von Stecher} J 2010 {\em J. Phys. B: At. Mol. Opt. Phys.\/} {\bf 43} 101002

\bibitem{Vonstecher2011fas}
von Stecher J 2011 {\em Phys. Rev. Lett.\/} {\bf 107} 200402

\bibitem{Deltuva2012ubt}
Deltuva A 2012 {\em Phys. Rev. A\/} {\bf 85} 012708

\bibitem{Ferlaino2009efu}
Ferlaino F, Knoop S, Berninger M, Harm W, {D'Incao} J~P, N\"agerl H~C and Grimm
  R 2009 {\em Phys. Rev. Lett.\/} {\bf 102} 140401

\bibitem{Mehta2009gtd}
Mehta N~P, Rittenhouse S~T, D'Incao J~P, von Stecher J and Greene C~H 2009 {\em
  Phys. Rev. Lett.\/} {\bf 103} 153201

\bibitem{Fano1976doe}
Fano U 1976 {\em Phys. Today\/} {\bf 29} 32

\bibitem{Berry1966scs}
Berry M~V 1966 {\em Proc. Phys. Soc. London\/} {\bf 88} 285

\bibitem{Berninger2011uot}
Berninger M, Zenesini A, Huang B, Harm W, N\"{a}gerl H~C, Ferlaino F, Grimm R,
  Julienne P~S and Hutson J~M 2011 {\em Phys. Rev. Lett.\/} {\bf 107} 120401

\bibitem{Berninger2012frw}
Berninger M, Zenesini A, Huang B, Harm W, N\"agerl H~C, Ferlaino F, Grimm R,
  Julienne P~S and Hutson J~M 2013 {\em Phys. Rev. A\/} {\bf 87} 032517

\bibitem{Weber2003tbr}
Weber T, Herbig J, Mark M, N\"agerl H~C and Grimm R 2003 {\em Phys. Rev.
  Lett.\/} {\bf 91} 123201

\bibitem{Braaten2001tbr}
Braaten E and Hammer H~W 2001 {\em Phys. Rev. Lett.\/} {\bf 87} 160407

\bibitem{Greene2004arf}
Greene C~H, Esry B~D and Suno H 2004 {\em Nucl. Phys. A\/} {\bf 737} 119--124

\bibitem{Bohn1998epf}
Bohn J~L, Esry B~D and Greene C~H 1998 {\em Phys. Rev. A\/} {\bf 58} 584--597

\end{thebibliography}

\providecommand{\newblock}{}

\end{document}